\documentclass[final]{svjour3}
\usepackage{graphicx}
\usepackage{rotating}
\usepackage{amssymb}
\usepackage{mathptmx}
\usepackage[numbers,sort&compress]{natbib}
\makeatletter
\journalname{Journal of Low Temperature Physics}
\bibpunct{[}{]}{,}{n}{}{,}

\begin{document}

\newcommand{\hdblarrow}{H\makebox[0.9ex][l]{$\downdownarrows$}-}
\title{SPT-SLIM: A Line Intensity Mapping Pathfinder for the South Pole Telescope}

\author{K.~S.~Karkare~$^{1,2}$ \and A.~J.~Anderson~$^2$ \and P.~S.~Barry~$^{3,1,4,5}$ \and B.~A.~Benson~$^{2,4,1}$ \and J.~E.~Carlstrom~$^{4,1,3}$ \and T.~Cecil~$^3$ \and C.~L.~Chang~$^{3,4,1}$ \and M.~A.~Dobbs~$^6$ \and M.~Hollister~$^2$ \and G.~K.~Keating~$^7$ \and D.~P.~Marrone~$^8$ \and J.~McMahon~$^{1,4,9}$ \and J.~Montgomery~$^6$ \and Z.~Pan~$^3$ \and G.~Robson~$^5$ \and M.~Rouble~$^6$ \and E.~Shirokoff~$^{1,4}$ \and G.~Smecher~$^{10}$}
\authorrunning{K.~S.~Karkare et al.}

\institute{\email{kkarkare@kicp.uchicago.edu}\\
1: Kavli Institute for Cosmological Physics, University of Chicago, Chicago, IL 60637, USA\\
2: Fermi National Accelerator Laboratory, Batavia, IL 60510, USA\\
3: High-Energy Physics Division, Argonne National Laboratory, Lemont, IL 60439, USA\\
4: Department of Astronomy and Astrophysics, University of Chicago, Chicago, IL 60637, USA\\
5: School of Physics \& Astronomy, Cardiff University, Cardiff CF24 3AA, UK\\
6: Department of Physics and McGill Space Institute, McGill University, Montreal, Quebec H3A2T8, Canada\\
7: Harvard-Smithsonian Center for Astrophysics, Cambridge, MA 02138, USA\\
8: Steward Observatory, University of Arizona, Tucson, AZ 85721, USA\\
9: Department of Physics, University of Chicago, Chicago. IL 60637, USA\\
10: Three-Speed Logic, Inc., Victoria, B.C., V8S 3Z5, Canada}

\maketitle

\begin{abstract}

The South Pole Telescope Summertime Line Intensity Mapper (SPT-SLIM) is a pathfinder experiment that will demonstrate the use of on-chip filter-bank spectrometers for mm-wave line intensity mapping (LIM). The SPT-SLIM focal plane consists of 18 dual-polarization $R=300$ filter-bank spectrometers covering 120--180 GHz, coupled to aluminum kinetic inductance detectors. A compact cryostat holds the detectors at 100 mK and performs observations without removing the SPT-3G receiver. SPT-SLIM will be deployed to the 10-m South Pole Telescope for observations during the 2023--24 austral summer. We discuss the overall instrument design, expected detector performance, and sensitivity to the LIM signal from CO at $0.5 < z < 2$. %, and projected constraints on cold molecular gas during the peak of cosmic star formation. 
The technology and observational techniques demonstrated by SPT-SLIM will enable next-generation LIM experiments that constrain cosmology beyond the redshift reach of galaxy surveys.


\end{abstract}

%When preparing your manuscript, please follow the instructions in this template. This template serves as a universal template for Special Issue Articles to be taken into consideration for publication in the Journal of Low Temperature Physics.\\
%Insert an empty line after the section title. Indent at the start of each paragraph after the first paragraph of the section, which is not indented. For special issue articles the page length is 6 pages. Invited contributions may be 12 pages long. In order to estimate your article length, please prepare your manuscript in Times or Times New Roman, font size 11, justify the body text, and make sure the page format is set to A4. The template margins are: Top: 2.07”, Left: 1.77”, Bottom: 2.17”, Right: 1.77” inches.\\
%Subsections can be used. 

%\newpage 

\section{Introduction}

Measurements of galaxies that trace large-scale structure (LSS)  constrain the physics of our cosmological model and the astrophysics of star and galaxy formation throughout the history of the Universe. However, large-scale optical galaxy surveys are limited to low redshifts where individual objects can be detected, and facilities that can detect high-redshift galaxies are limited to small survey areas. To make progress on open questions in fundamental cosmology and early-Universe astrophysics, we are motivated to extend the redshift reach and cosmic volume accessible by LSS surveys.

%The $\Lambda$CDM cosmological model---anchored by observations of the cosmic microwave background (CMB), large-scale structure (LSS), and Type 1a supernovae---describes the statistical properties of the Universe with only six free parameters. But while increasingly precise measurements have determined some parameters to sub-percent levels, we still lack an understanding of the fundamental nature of dark energy, dark matter, and inflation. At the same time, the properties of the first stars and galaxies, and the dynamics of the Epoch of Reionization, remain largely unobserved. To make progress on open questions in cosmology and early-Universe astrophysics, we are motivated to extend the redshift reach and cosmic volume accessible by LSS surveys.

Line intensity mapping (LIM) uses low angular resolution spectroscopic observations of emission from atomic or molecular lines to map LSS in three dimensions \cite{kovetz17}. At high redshift, this allows large cosmological volumes to be measured more efficiently than with galaxy surveys. LIM targeting far-IR emission lines (e.g., CO rotational lines or the [CII] ionized carbon fine structure line) is particularly promising for measuring high-redshift LSS, and is capable of providing competitive constraints on the expansion history \cite{karkare18}, neutrinos \cite{moradinezhad21}, and the dynamics of the reionization process \cite{gong12}. But while first detections of the mm-wave LIM signal are now being reported \cite{keating20}, current-generation instruments lack the sensitivity for competitive constraints on cosmology and high-redshift astrophysics. New technology is needed to build significantly more sensitive instruments.

In this paper we present the South Pole Telescope Summertime Line Intensity Mapper (SPT-SLIM), a pathfinder demonstrating the critical, enabling technology for mm-wave LIM: on-chip, large-format mm-wave spectrometers. On-chip spectrometers (e.g., SuperSpec \cite{shirokoff12}, DESHIMA \cite{endo19}, $\mu$Spec \cite{cataldo18}) pack an order of magnitude more detectors in a limited cold volume compared to current instruments \cite{crites14, cothard20}. Building on CMB-heritage techniques in detector fabrication and readout, this technology promises to rapidly improve the instantaneous sensitivity of mm-wave spectrometers and enable deep measurements of LSS over a wide redshift range. SPT-SLIM will deploy to the South Pole in the 2023--2024 austral summer, observing with the highest-density on-chip mm-wave spectrometer to date, and will be the first ground-based experiment to perform LIM with this technology. To accommodate the large detector counts needed for spectroscopy, we will use aluminum (Al) kinetic inductance detectors (KIDs) and next-generation frequency-domain multiplexed RF readout. 

This paper is organized as follows: in Section~\ref{sec:LIM} we discuss SPT-SLIM's mm-wave LIM measurement, in Section~\ref{sec:cryostat} we present the cryostat design, in Section~\ref{sec:detectors} we discuss the focal plane, detectors, and readout, and in Section~\ref{sec:deployment} we present projected constraints on the LIM power spectrum and our deployment schedule. Two other manuscripts in these proceedings discuss the SPT-SLIM filter-bank simulations (Robson et al. \cite{robson21}) and the focal plane design (Barry et al. \cite{barry21}).

\section{Millimeter-Wave Line Intensity Mapping with SPT-SLIM}
\label{sec:LIM}

LIM measurements detect large-scale fluctuations in atomic or molecular line emission from many unresolved sources. Using a moderate-resolution spectrometer and knowledge of the target line's rest-frame wavelength, a 3-dimensional spectral-spatial data cube of line emission can be measured. The power spectrum of the surveyed volume is sensitive to a combination of the line emission physics and the underlying dark matter distribution, which probes the cosmology.

SPT-SLIM will observe from 120--180 GHz, targeting CO $\mathrm{J} \rightarrow \mathrm{J}-1$ rotational transitions from high-redshift galaxies with $R=300$ spectroscopy and arcminute angular resolution. These rotational lines originate in dense molecular clouds and trace the sites of star formation. In the 2 mm band we will detect CO(2-1) from $0.3 < z < 0.9$, CO(3-2) from $0.9 < z < 1.8$, and CO(4-3) from $ 1.6 < z < 2.8$.  By measuring the amplitudes of multiple CO lines at different times in cosmic history, SPT-SLIM will inform models of the CO luminosity function and the cold molecular gas fraction as a function of redshift for the full population of galaxies. 

\begin{figure}[htbp]
\begin{center}
\includegraphics[width=1.0\linewidth, keepaspectratio]{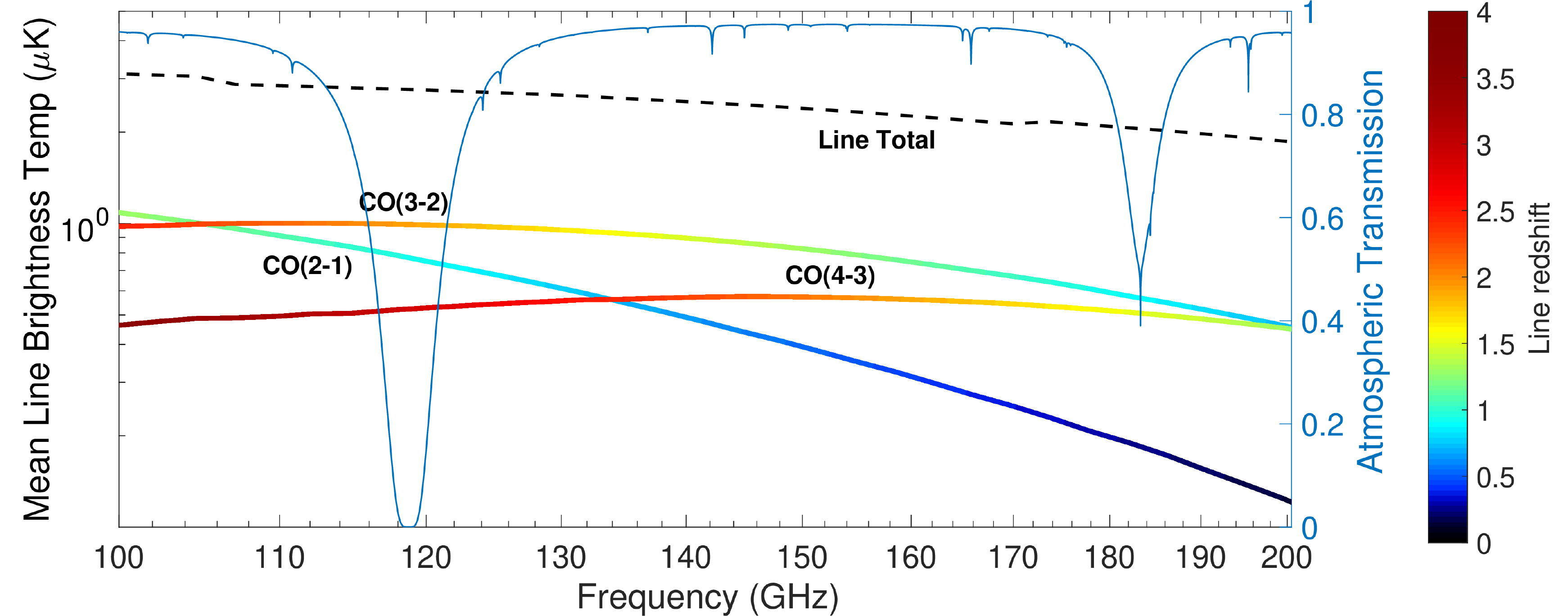}
\caption{Spectral lines targeted by SPT-SLIM, which will observe from 120--180 GHz. Shown are the model predictions for the brightness temperatures of CO(2-1), CO(3-2), and CO(4-3), their sum total (dashed black), and median atmospheric transmission during the South Pole summer (thin blue). The colors correspond to the observed redshifts of the lines. (Color figure online.)}
\end{center}
\label{fig:line_temps}
\vspace{-15pt}
\end{figure}

Fig.~1 shows the expected CO brightness temperature as a function of observing frequency, in addition to the median atmospheric transmission at the South Pole during the austral summer calculated with \textit{am} \cite{am}. Line confusion---i.e., emission from multiple redshifts at the same observing frequency---will present a challenge when analyzing the SPT-SLIM dataset, as seen in Fig.~1. Many approaches to mitigating line confusion exist in the literature, including masking galaxies with known redshifts, using the power spectrum anisotropy in redshift space, and cross-correlating within a dataset or with external galaxy samples \cite{sun18, lidz16, cheng20}. We anticipate using internal cross-spectra across the 60 GHz bandwidth and cross-correlations with external galaxy catalogs to isolate emission at each redshift.

%\begin{figure}[htbp]
%\begin{center}
%\includegraphics[width=0.8\linewidth, keepaspectratio]{fig1.eps}
%\caption{Insert the figure caption directly after the header. 
%Use the words {\it Top, Bottom, Left, } and {\it Right} (in italics) 
%to denote the sub figures being described. 
%Additional annotation should also be in italics, 
%e.g., {\it solid symbols, open symbols, red, dashed line,} etc. If including color figures, please supply a %B\&W or greyscale figure as well for print and include the following statement at the end of your caption: %(Color figure online.)}
%\end{center}
%\label{fig1}
%\end{figure}

%\subsection{Equations}

%Place equations in the text and number them, for example:
%\begin{equation}
%	\frac{\rho-\rho_{0}}{\rho_{0}} = aH + bH^{2},
%\end{equation}
%When referring to equations please use parentheses and the number of the equation as follows: Eq. (1).

%\subsection{Figures}

%When inserting figures, Leave an empty line after the Figure caption. When referring to figures in the text, please do so as follows: Fig.~\ref{fig1}.

\section{Cryostat}
\label{sec:cryostat}

SPT-SLIM will use a compact, custom cryostat that is mounted in the SPT receiver cabin without needing to remove the SPT-3G receiver. In this section we discuss the design of the receiver cryostat and optics.

%\subsection{Receiver Design and Cryogenics}

\textbf{Receiver Design and Cryogenics}: The SPT is a 10m aperture off-axis Gregorian telescope, configured such that the receiver cabin does not obstruct the beam \cite{carlstrom11}. The receiver cabin houses the SPT-3G receiver; a powered secondary and flat tertiary mirror couple this cryostat to the main dish \cite{sobrin21}. As illustrated in Fig.~2 left, there is additional room in the cabin behind the SPT-3G tertiary, towards the primary. The Event Horizon Telescope (EHT) has previously used this space to install a cryostat and conduct observations; SPT-SLIM will be positioned in the same location as the EHT receiver \cite{kim18}. We have designed a cryostat that can be lowered into place from the roof of the receiver cabin when docked. A Cryomech PT407 pulse tube cooler mounted to the side of the cryostat will provide cooling to 4K. To maximize the sensitivity of the Al KIDs, the focal plane array must be held at 100 mK. A High Precision Devices 155 adiabatic demagnetization refrigerator (ADR) will provide the sub-Kelvin cooling; with the expected loading (dominated by the coaxial lines used to read out the detectors), we anticipate a duty cycle of $\sim 70\%$.  The internal cryostat components are shown in Fig.~2 right.

\begin{figure}[htbp]
\begin{center}
\includegraphics[width=1.0\linewidth, keepaspectratio]{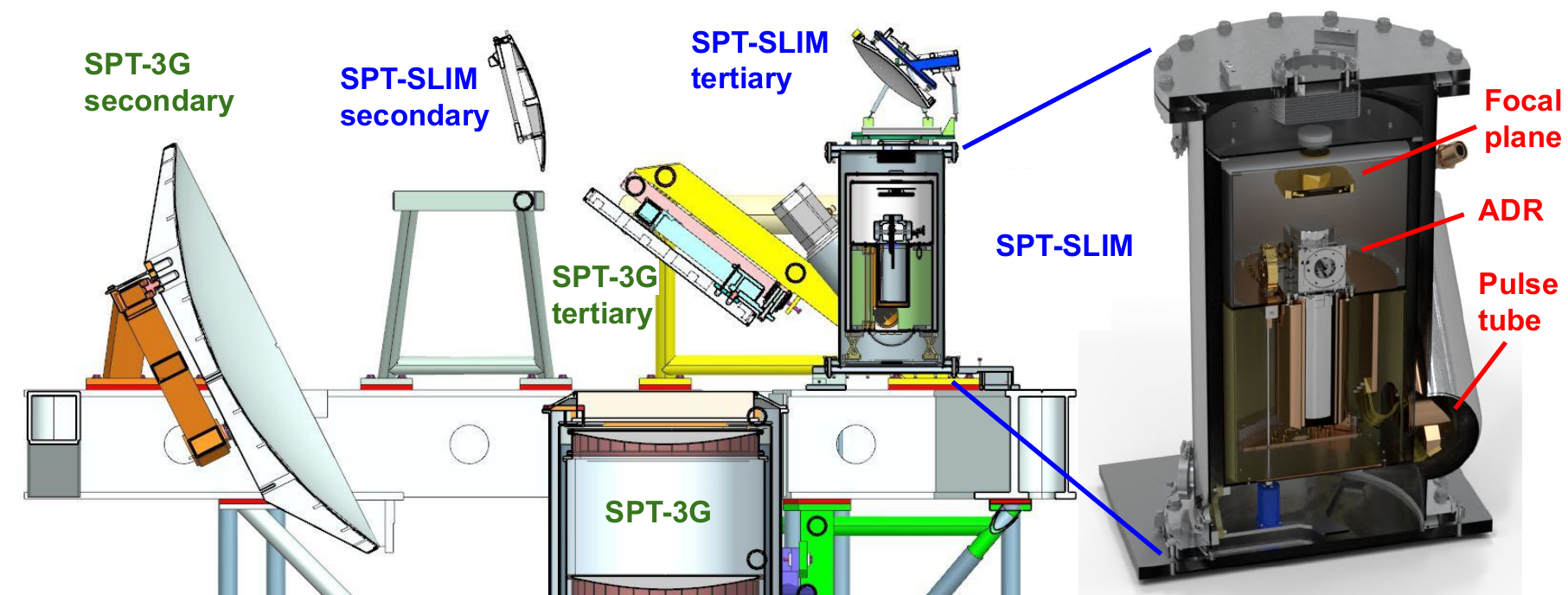}
\caption{The SPT-SLIM cryostat and mounting.  \textit{Left}: The SPT-SLIM cryostat mounts in the SPT receiver cabin without removing the SPT-3G receiver, and uses two auxiliary mirrors to redirect the beam. \textit{Right}: The cryostat uses an adiabatic demagnetization refrigerator to cool the focal plane to 100 mK, backed by a PT407 pulse tube. The focal plane support structure is not shown. (Color figure online.)}
\end{center}
\label{fig:cryostat}
\vspace{-15pt}
\end{figure}

%\subsection{Optical Design}

\textbf{Optical Design}: SPT-SLIM adapts the EHT optical design to redirect the beam that normally illuminates the SPT-3G receiver. Two custom mirrors---a hyperbolic secondary and ellipsoidal tertiary---will pick off the beam and focus it into the SPT-SLIM cryostat. We plan to reuse the EHT secondary mirror, and have designed a new tertiary mirror to accommodate the wider field-of-view required for the SPT-SLIM focal plane. Removing these mirrors and installing the cryostat is straightforward when the telescope is docked. The $f/2.2$ optics provide a 40mm diameter diffraction-limited area at the focal plane, allowing for 18 spatial pixels with a pixel spacing of 1.5--2 F$\lambda$. The HDPE entrance window has an aperture of 100 mm. Infrared filtering consists of a stack of Zotefoam filters at 300 K, HDPE and nylon filters at 50 K, and a nylon and metal-mesh filter (7 icm cutoff) at 4 K. The internal cryostat optics are unpowered and anti-reflection coated with expanded Teflon. 

\section{Detectors and Readout}
\label{sec:detectors}

SPT-SLIM will deploy an array of 18 dual-polarization $R=300$ on-chip filter-bank spectrometers. Here we discuss the design of the focal plane, filter-bank, detectors, and readout; see the additional companion proceedings for more details \cite{robson21, barry21}.

%\subsection{Focal Plane Array}

\textbf{Focal Plane Array}: The SPT-SLIM focal plane is a hexagonal close-packed array, where each of the 18 spatial pixels is sensitive to both polarizations (36 spectrometers total). As seen in Fig.~3 left and middle, radiation is admitted by conical profiled horns made from gold-plated, machined Al and coupled by a planar orthomode transducer (OMT) to separate pairs of superconducting microstrip lines, which are combined with a 180$^{\circ}$ hybrid. This design covers the full 120--180 GHz band with high efficiency. The focal plane consists of three identical submodules. Each submodule contains 6 dual-polarization pixels, coupled to 12 spectrometers. The length of each filter-bank is 80 mm, oriented radially from the focal plane center. Dowel pins and deep reactive ion etch features align the three submodules to a common interface plate attached to the horn block. %
%
%The backshort wafer, which contains a mm-wave choke to reduce inter-pixel leakage, is formed from a DRIE etched and metallized Si wafer. 
The Si wafer stack is then secured with BeCu spring clamps to minimize microphonic noise. We have successfully demonstrated this mechanical process with OMT-KID prototype devices \cite{tang20}.

\begin{figure}[htbp]
\begin{center}
\includegraphics[width=1.0\linewidth, keepaspectratio]{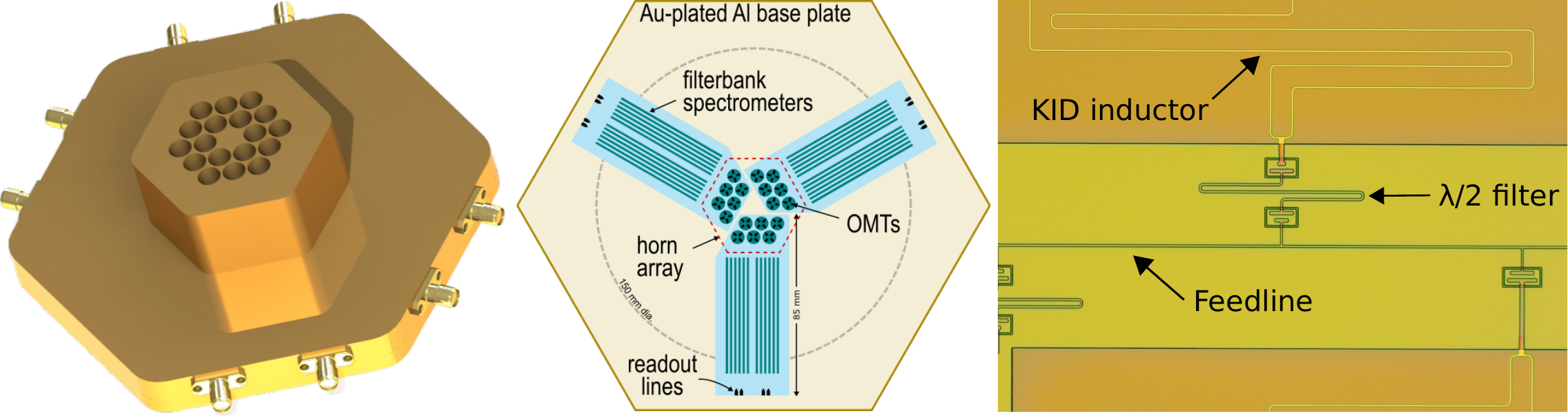}
\caption{The SPT-SLIM horn block, focal plane, and filter-bank. \textit{Left}: The focal plane packaging consists of the horn block and coaxial lines for readout. \textit{Middle}: Schematic of the focal plane with 18 dual-polarization OMTs feeding the filter-banks. \textit{Right}: Microscope image of the feedline and a capacitively-coupled $\lambda/2$ filter, which is coupled to the KID inductor.}
\end{center}
\label{fig:fpu}
\vspace{-15pt}
\end{figure}

%\subsection{Filter-Bank Design}

\textbf{Filter-Bank Design}: Each spatial pixel is coupled to an $R= \lambda / \Delta \lambda = 300$ superconducting filter-bank spectrometer, broadly based on the SuperSpec design \cite{redford18, karkare20}. Broadband radiation is split into spectral channels via a bank of filter elements composed of capacitively-coupled $\lambda / 2$ mm-wave resonators (Fig.~3 right). The center frequency of each channel is set by the resonator length; the resolution is controlled by the coupling strength to each filter element and is ultimately limited by dielectric loss. Filters are spaced so that the Lorentzian profiles overlap to provide a Nyquist-sampled spectrum (oversampling factor of 2). Each filter-bank consists of 200 resonators spanning 120--180 GHz. From simulations that incorporate realistic fabrication errors and dielectric loss tangents we expect that a total filter-bank efficiency of $\sim 60\%$ is achievable with this design.

%\subsection{KID Design}

\textbf{KID Design}: The SPT-SLIM detectors are microstrip coupled lumped-element KIDs~\cite{tang20}, where the resonator is formed from a discrete inductor and capacitor. At mm wavelengths, the inductor also doubles as a lossy transmission line coupled to the output of each filter. Absorbing the incident radiation as it propagates along the KID, the inductor must remain a superconducting high-$Q$ material at the KID resonant frequency, which constrains the superconductor to $T_c < 2$ K. We will use a hybrid Al-Nb resonator, where the inductor/absorber is Al ($T_c \sim 1.3$ K) and the capacitor is Nb. This design makes use of well-characterized properties of Al for applications operating at similar sensitivities, and a Nb-Si interface that has shown reduced two-level system noise. 
%PETE DESCRIPTION OF COUPLING SCHEME. 

Based on typical conditions at the SPT, for an $R=300$ spectral channel we expect a loading of around 25 fW, and a noise target of $\sim 3 \times 10^{-18}$ W Hz$^{-1/2}$. This per-detector sensitivity has been demonstrated on large-scale Al-based KID arrays \cite{baselmans17}. To minimize readout complexity and achieve a multiplexing density of at least 2k detectors per line, we require tight control over resonator placement. We have demonstrated a process that enables near-perfect detector yield by lithographically modifying detectors after testing, changing the resonant frequencies to separate resonators that were originally collided \cite{mcgeehan18}. With a measured fractional accuracy of $< 2 \times 10^{-5}$, all intact resonators will be recoverable.

%\subsection{Readout}

%MACLEAN's paragraph on readout description, total channel count, etc. 
\textbf{Readout}: The detectors will be read out with the ICE signal processing platform \cite{bandura16} that has been adapted to a new operational domain with GHz KIDs. The ICE system has been successfully deployed in a variety of experimental contexts, including SPT-3G and CHIME. The system consists of a general-purpose Xilinx Kintex-7 FPGA motherboard (the ``Iceboard''), which couples to modular daughterboards specifically designed or chosen for each experimental use-case, which house the analog-to-digital and digital-to-analog converters. For SPT-SLIM each Iceboard will be equipped with a new frontend, consisting of two RF chains that can each read out up to 1024 frequency-domain multiplexed detector channels over 500~MHz of bandwidth. We will use 5 Iceboards to read out SPT-SLIM's 8640 KIDs. The tones can be placed anywhere from 0--6~GHz through direct digitization without external IQ mixing; each tone operates independently. To maintain each detector's linearity over a range of loading conditions, tone tracking may be implemented either continuously, using the existing digital-active nulling feedback, or occasionally by updating each tone when the resonant frequency of the detector shifts by a substantial fraction of the resonator linewidth. %The demonstrated performance and maturity of the ICE system (SPT-3G has continuously operated 32 Iceboards since 2017, achieving background-limited white noise performance, excellent low-frequency stability, and negligible readout-related downtime) provide the basis for the expansion of its associated firmware, software, and hardware into this novel operational domain.

\section{Observing Plan and Projected Sensitivity}
\label{sec:deployment}

SPT-SLIM will deploy to the South Pole during the 2023--2024 austral summer season. After SPT-3G has completed operations for the year, we will install SPT-SLIM and its mirrors in the receiver cabin, commission and calibrate the instrument, and begin science observations in January 2024. We anticipate observing for up to four weeks, targeting an approximately $0.1^{\circ} \times 10^{\circ}$ patch using constant-elevation scans. The receiver will then be removed in time for SPT-3G winter operations in February 2024. 

In one observing season, SPT-SLIM will have the sensitivity to make robust detections of several CO lines in both the clustering and shot noise regimes. Fig.~4 shows projected constraints on the CO(2-1), CO(3-2), and CO(4-3) power spectra at the redshifts corresponding to the 120--180 GHz observing band. These projections assume an end-to-end optical efficiency of 25\% and detector loading dominated by the South Pole atmosphere during median Dec-Jan-Feb conditions; see Fig.~1. Three curves are shown corresponding to integration times of 100, 200, and 300 hours, bracketing pessimistic to optimistic observing times obtainable in a month of deployment. For 200 hrs, we project 7, 19, and 13$\sigma$ detections of the total power spectra for $\mathrm{J}_{\mathrm{upper}} = 2,3,4$ respectively. With these measurements, SPT-SLIM will discriminate between several models of CO luminosity functions in the literature, particularly in the range of $L'_{\rm CO} = 10^{10}$--$10^{11}$ K km/s pc$^{-2}$.

\begin{figure}[htbp]
\begin{center}
\includegraphics[width=1.0\linewidth, keepaspectratio]{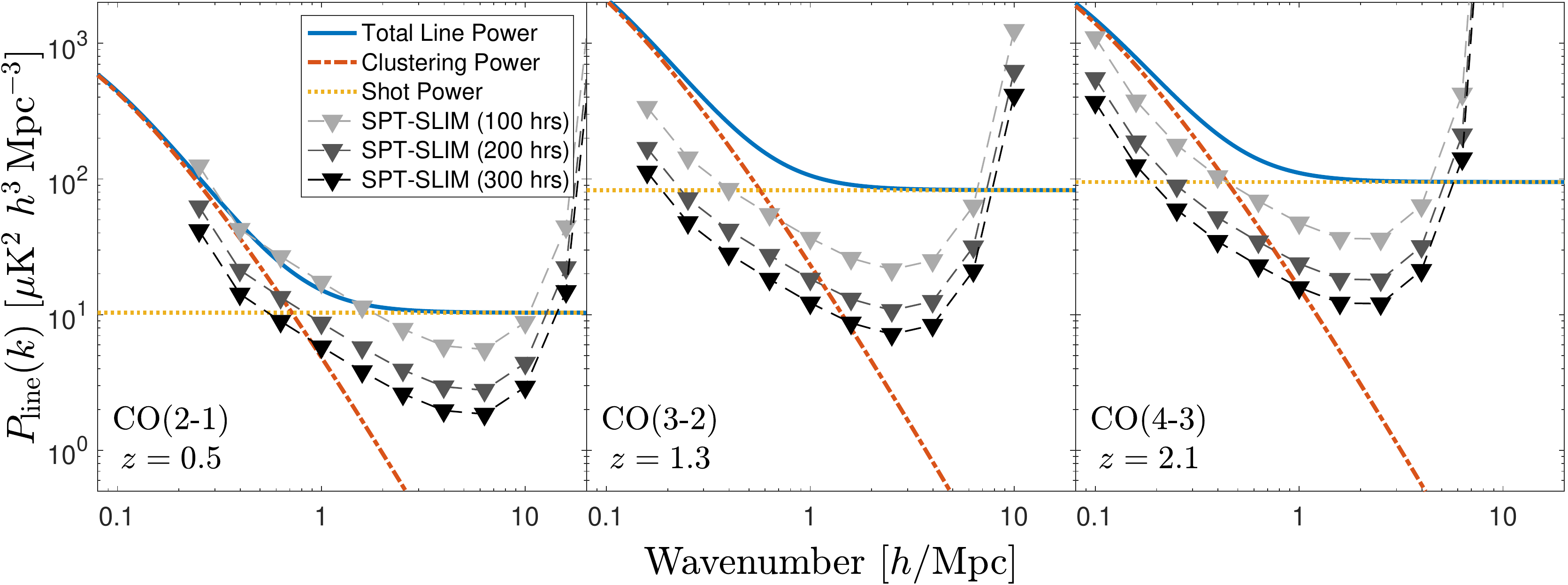}
\caption{SPT-SLIM's projected constraints on  CO power spectra at $z=0.5$ (\textit{left}), $z=1.3$ (\textit{middle}), and $z=2.1$ (\textit{right}) in the 120--180 GHz band. The expected clustering power spectra are in orange, the shot noise power spectra are in yellow, and the totals are in blue. Anticipated noise curves for 100, 200, and 300 hours of integration are shown with gray and black arrows. (Color figure online.)}
\end{center}
\label{fig:powspec}
\vspace{-15pt}
\end{figure}

% IF ROOM, include luminosity function image

While SPT-SLIM is primarily a LIM pathfinder, the SPT's arcminute resolution also makes it capable of pointed observations and redshift identification of dusty star-forming galaxies (DSFGs). In one hour of observing a typical $10^{13}$ L$_{\odot}$ DSFG, SPT-SLIM will detect CO(2-1) from $z=1.5$ at $7\sigma$, or CO(4-3) from $z=3$ at $4 \sigma$. In just a few days, SPT-SLIM could determine redshifts for hundreds of DSFGs identified in SPT broadband surveys, which could then be followed up at high resolution with ALMA.

In future seasons SPT-SLIM will continue to serve as a pathfinder. Next-generation mm-wave LIM science will benefit from wide-bandwidth observations, and SPT-SLIM could be outfitted with spectrometers in the 80--120 GHz and 180--310 GHz bands. %The cryostat could also be installed during the Austral winter season for deep observations, as done with EHT. 
After demonstration with SPT-SLIM, compact wide-bandwidth spectrometers will be promising candidates for next-generation receivers on the SPT or other facilities with the sensitivity to constrain cosmology and astrophysics beyond the reach of galaxy surveys.

\begin{acknowledgements}
%(Optional) Please enter text directly after the header. Template v.1 by KLL - June 18, 2015.
This work is supported by Fermilab under award LDRD-2021-048 and by the National Science Foundation under award AST-2108763. K.~S.~Karkare is supported by an NSF Astronomy and Astrophysics Postdoctoral Fellowship under award AST-2001802.
\end{acknowledgements}

%\pagebreak

%Regarding references: please see the example bibliography below. Please include DOI numbers wherever possible. Please number your references in order of appearance in the text. Citations should be with square brackets as follows [1], [2,3] or [4-7]. 
%Please use the commonly used and accepted journal abbreviation whenever possible. For example, the Journal of Low Temperature Physics is J. Low Temp. Phys. \\
%When citing articles that shall appear in the same special issue, please do the following. During the review stage you will be asked to cite the article as A. Name, J. Low Temp. Phys. This Special Issue (YEAR). Upon acceptance of the article and receiving the proofs, you will be requested to update the references and replace ''This Special Issue'' with the full citation and DOI number. Articles that have been accepted and published online (on SpringerLink) will immediately have a DOI number even if they are not assigned to an issue yet. When citing articles that appeared in previous special issues of this journal, please use the standard citation and include the DOI number.


\begin{thebibliography}{99}

\bibitem{kovetz17} E. Kovetz et al., {\it arXiv eprints}:1709.09066
  (2017)

\bibitem{karkare18} K. S. Karkare et al., {\it Phys. Rev. D.}
  \textbf{98}, 043529 (2018). DOI: 10.1103/PhysRevD.98.043529
  
\bibitem{moradinezhad21} A. Moradinezhad Dizgah et al., {\it arXiv eprints}:2110.00014 (2021)

\bibitem{gong12} Y. Gong et al., {\it Astroph. J.} \textbf{745}, 49 (2012). DOI: 10.1088/0004-637X/745/1/49

\bibitem{keating20} G. Keating et al., {\it Astroph. J.} \textbf{901}, 141 (2020). DOI: 10.3847/1538-4357/abb08e

\bibitem{shirokoff12} E. Shirokoff et al., {\it Proc. SPIE} \textbf{8452}, 84520R (2012). DOI: 10.1117/12.927070
  
\bibitem{endo19} A. Endo et al., {\it J. Astr. Tel. Instr. Syst.}
  \textbf{5}(3), 035004 (2019). DOI: 10.1117/1.JATIS.5.3.035004

\bibitem{cataldo18} G. Cataldo et al., {\it J. Low Temp. Phys.}
  \textbf{193}, 923 (2018). DOI: 10.1007/s10909-018-1902-7

\bibitem{crites14} A. T. Crites et al., {\it Proc. SPIE}
  \textbf{9153}, 91531W (2014). DOI: 10.1117/12.2057207

\bibitem{cothard20} N. F. Cothard et al., {\it J. Low Temp. Phys.} \textbf{199}, 878 (2020). DOI: 10.1007/s10909-019-02297-1

\bibitem{robson21} G. Robson et al., {\it J. Low Temp. Phys.} This Special Issue (2021)

\bibitem{barry21} P. Barry et al., {\it J. Low Temp. Phys.} This Special Issue (2021)
  
\bibitem{am} S. Paine, {\it The am atmospheric model},
https://doi.org/10.5281/zenodo.1193646 (2018)

\bibitem{sun18} G. Sun et al., {\it Astrophys. J.} \textbf{856}, 107 (2018). DOI: 10.3847/1538-4357/aab3e3

\bibitem{lidz16} A. Lidz et al., {\it Astrophys. J.} \textbf{825}, 143 (2016). DOI: 10.3847/0004-637X/825/2/143

\bibitem{cheng20} Y. T. Cheng et al., {\it Astrophys. J.} \textbf{901}, 142 (2020). DOI: 10.3847/1538-4357/abb023

\bibitem{carlstrom11} J. E. Carlstrom et al., {\it Publ. Astron. Soc. Pac.} \textbf{123}, 568 (2011). DOI: 10.1086/659879

\bibitem{sobrin21} J. A. Sobrin et al., {\it arXiv eprints}:2106.11202 (2021)

\bibitem{kim18} J. Kim et al., {\it Proc. SPIE} \textbf{10708}, 107082S (2018). DOI: 10.1117/12.2301005

\bibitem{redford18} J. Redford et al., {\it Proc. SPIE} \textbf{10708},
  107081O (2018). DOI: 10.1117/12.2313666
  
\bibitem{karkare20} K. S. Karkare et al., {\it J. Low Temp. Phys.} \textbf{199}, 849 (2020). DOI: 10.1007/s10909-020-02407-4

\bibitem{tang20} Q. Y. Tang et al., {\it J. Low Temp. Phys.} \textbf{199}, 362 (2020). DOI: 10.1007/s10909-020-02341-5

\bibitem{baselmans17} J. Baselmans et al., {\it Astron. \& Astrophy.} \textbf{601}, A89 (2017). DOI: 10.1051/0004-6361/201629653 

\bibitem{mcgeehan18} R. McGeehan et al., {\it J. Low Temp. Phys.}
  \textbf{193}, 1024 (2018). DOI: 10.1007/s10909-018-2061-6
  
\bibitem{bandura16} K. Bandura et al., {\it J. Astr. Instru.} \textbf{5}(4), 1641005 (2016). DOI: 10.1142/S2251171716410051

\end{thebibliography}
\end{document}